\documentstyle[aps,multicol,prl,epsf,eqsecnum]{revtex}
\begin{document}

\wideabs{
\title{Energetics and Vibrational States for Hydrogen on Pt(111)}
\author{ S. C. Badescu$^{1,2}$, P. Salo$^{1}$, T. Ala-Nissila$^{1,2}$,
         S. C. Ying$^{2}$, K. Jacobi$^{3}$, Y. Wang$^{3}$,
         K. Bed\"{u}rftig$^{3}$, and G. Ertl$^{3}$ \\
$^{1}$\it{Helsinki Institute of Physics and Laboratory of Physics,
          Helsinki University of Technology,
          P.O. Box 1100, FIN-02015 HUT, Espoo, Finland}\\
$^{2}$\it{Department of Physics, Brown University, Providence,
          Rhode Island 02912-1843}\\
$^{3}$\it{Fritz-Haber-Institut der Max-Planck-Gesellschaft,
Faradayweg 4-6,
          D-14195 Berlin, Germany}}
\date{September 28, 2001}
\maketitle

\begin{abstract}

We present a combination of theoretical calculations and
experiments for the low-lying vibrational excitations of H and D
atoms adsorbed on the Pt(111) surface. The vibrational band states
are calculated based on the full three-dimensional adiabatic
potential energy surface obtained from first-principles
calculations. For coverages less than three quarters of a
monolayer, the observed experimental high-resolution electron
peaks at 31 and 68~meV are in excellent agreement with the
theoretical transitions between selected bands. Our results
convincingly demonstrate the need to go beyond the local harmonic
oscillator picture to understand the dynamics of this system.

\end{abstract}

}

Hydrogen on metal surfaces, and in particular on Pt(111) has
received considerable experimental
\cite{Hor99,DiW92,Chr76,Bar79,Ric87,Gra99,Jac01} and theoretical
\cite{Fei87,Fei97,Pap00,Ols99,Wat01,Nob00,XXu99,Kal02} attention.
Together with Pd, Pt is the most important material for
heterogenous catalysis of hydrogenation reactions. On the
fundamental side, hydrogen on metal surfaces provides a unique
opportunity to observe the crossover from classical to quantum
dynamics at relatively elevated  temperatures
\cite{Chr79,Pus85,Bad01,Cao97}.

 For the vibrational excitations of H/Pt(111), Bar\'{o} {\it et al.}
\cite{Bar79} applied high\textbf{-}resolution electron
energy\textbf{-}loss spectroscopy (HREELS) and observed for H~(D)
at coverages up to 0.7 monolayer (ML) two peaks at $68~(50)$~meV
(assigned to the symmetric stretch mode SS) and $153~(112)$~meV
(assigned to the asymmetric stretch mode AS). On the other hand,
Richter and Ho \cite{Ric87} found three peaks at $67~(51)$~meV,
$112~(84)$~meV and $153~(108)$~meV in their HREELS measurements
for a full ML of H~(D). They assigned these to the AS mode, the SS
mode, and  the unresolved overtone and combination losses,
respectively. To date, the disagreement between these observations
is not resolved. In addition, recent measurement of the diffusion
barrier for this system \cite{Gra99} yielded a very low value of
about $70$~meV, considerably less than a previous value of about
$200$~meV. All these discrepancies emphasize our lack of
understanding of this important system.

On the theoretical side, an earlier first-principles (FP) study
\cite{Fei87} at a full ML coverage has shown that H adsorbs in the
three-fold hollow fcc site. More recent FP calculations
\cite{Pap00,Ols99,Wat01,Nob00} at lower coverages indicate that
unlike for
 H/Ni(111) and H/Pd(111) \cite{Don98,Kre00,Lov98,Pau96}, the top site
 for H/Pt(111) is
also a local minimum only slightly higher in energy than the fcc
site. In addition,  the energy barrier for the fcc-hcp-fcc path is
only about $70$~meV.  In these calculations the main focus has
been on  adiabatic potential energy surface (APES). The curvatures
determined at the potential minima \cite{Fei87,Pap00,Ols99}
could not account quantitatively for the observed vibrational
excitation energies. Given the recent results of the shallow
barrier between the fcc and hcp sites, it is clear that
 quantum effects are essential in understanding the dynamics of this
 system, and  the vibrational spectroscopy need to be interpreted in terms of
excitations between vibrational band states as proposed earlier
\cite{Chr79,Pus85}.

In this Letter we investigate the energetics and vibrational
properties of H and D adatoms on the Pt(111) surface. The
experiments consist of HREELS investigations of H and D adsorption
on Pt(111) and its dependence on coverage. The experiments were
performed in an ultra-high vacuum (UHV) apparatus with a base
pressure of $3\times 10^{-11}$ mbar as described elsewhere
including sample preparation \cite{Bed99}. The HREEL spectrometer
was of the latest commercial Ibach design (Delta 0.5, SPECS,
Germany). The H$_2$ exposure was performed by backfilling the
preparation chamber while the sample was cooled to $85$ K with
liquid N$_2$. The exposure was varied between $0.1$ and $500$ L
($1\ {\rm L} = 1.33 \times 10^{-6}$~mbar$\times$s). The coverage
$\theta$ was determined from thermal desorption (TD) assuming
$\theta =1$ at saturation. The TD spectra of Christmann {\it et
al.} \cite{Chr76} were well reproduced. We focus on the spectrum
for the low coverage interval $\theta~\leq~0.75$ ML, within which
the HREEL spectra are qualitatively the same  \cite{peaks}.

\begin{figure}[htbp]
\unitlength1cm
\begin{center}
\begin{picture}(8.0,7.0)
\includegraphics{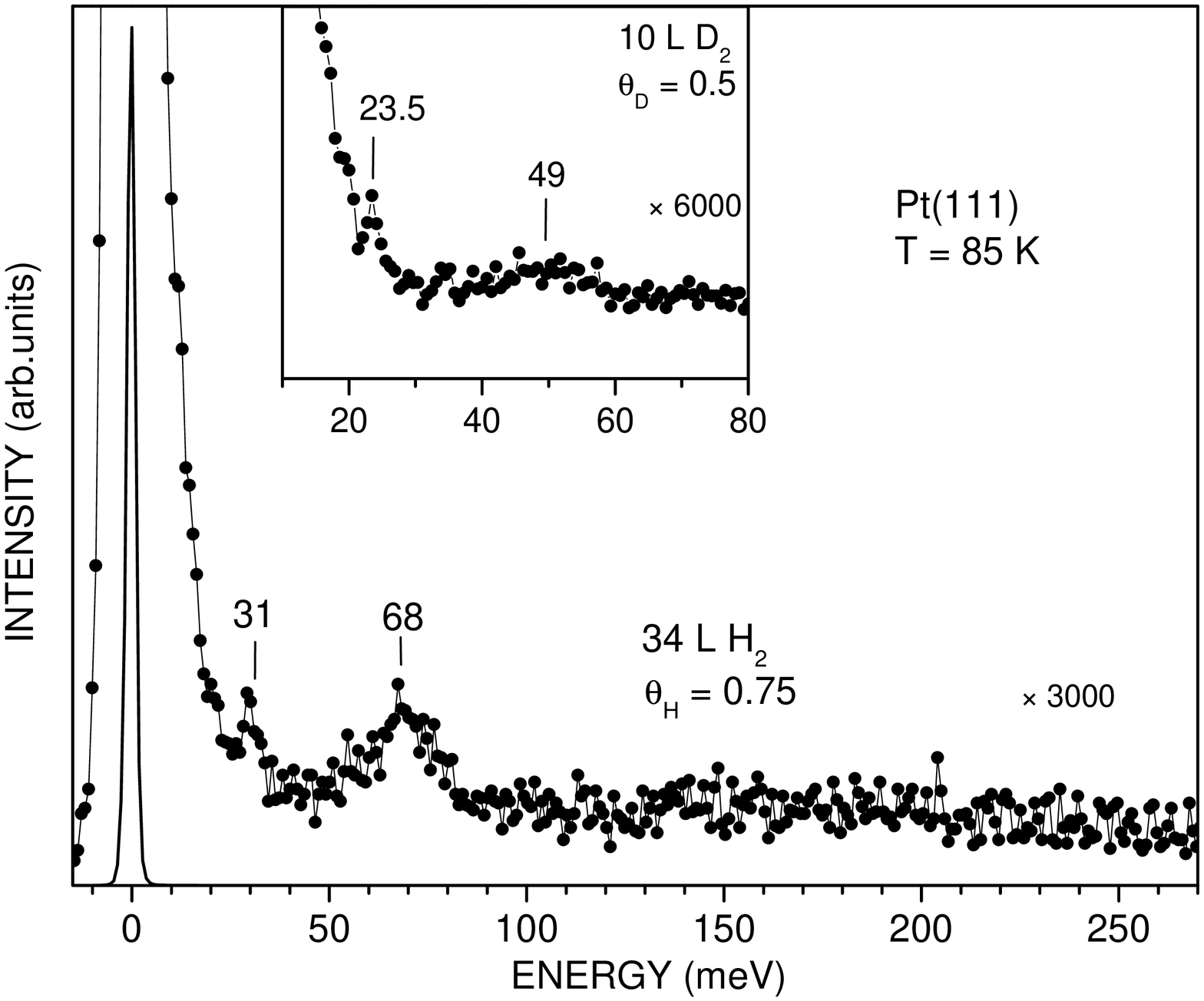}
\end{picture}
\caption{HREEL spectra for H on Pt(111) at $\theta$=$0.75$ ML(H)
and $0.5$ ML(D) for specular geometry. The spectrum for D is in the
inset, with the low-energy features displayed on a smaller
scale. The primary electron energy is E$_0$=2.0~eV. The exposure is given
in Langmuir (L) (1~L $\approx~1.33\times 10^{-6}$~mbar$\times$s).}
\label{HREELS1}
\end{center}
\end{figure}

\begin{figure}[htbp]
\unitlength1cm
\begin{center}
\begin{picture}(8.0,7.)
\includegraphics{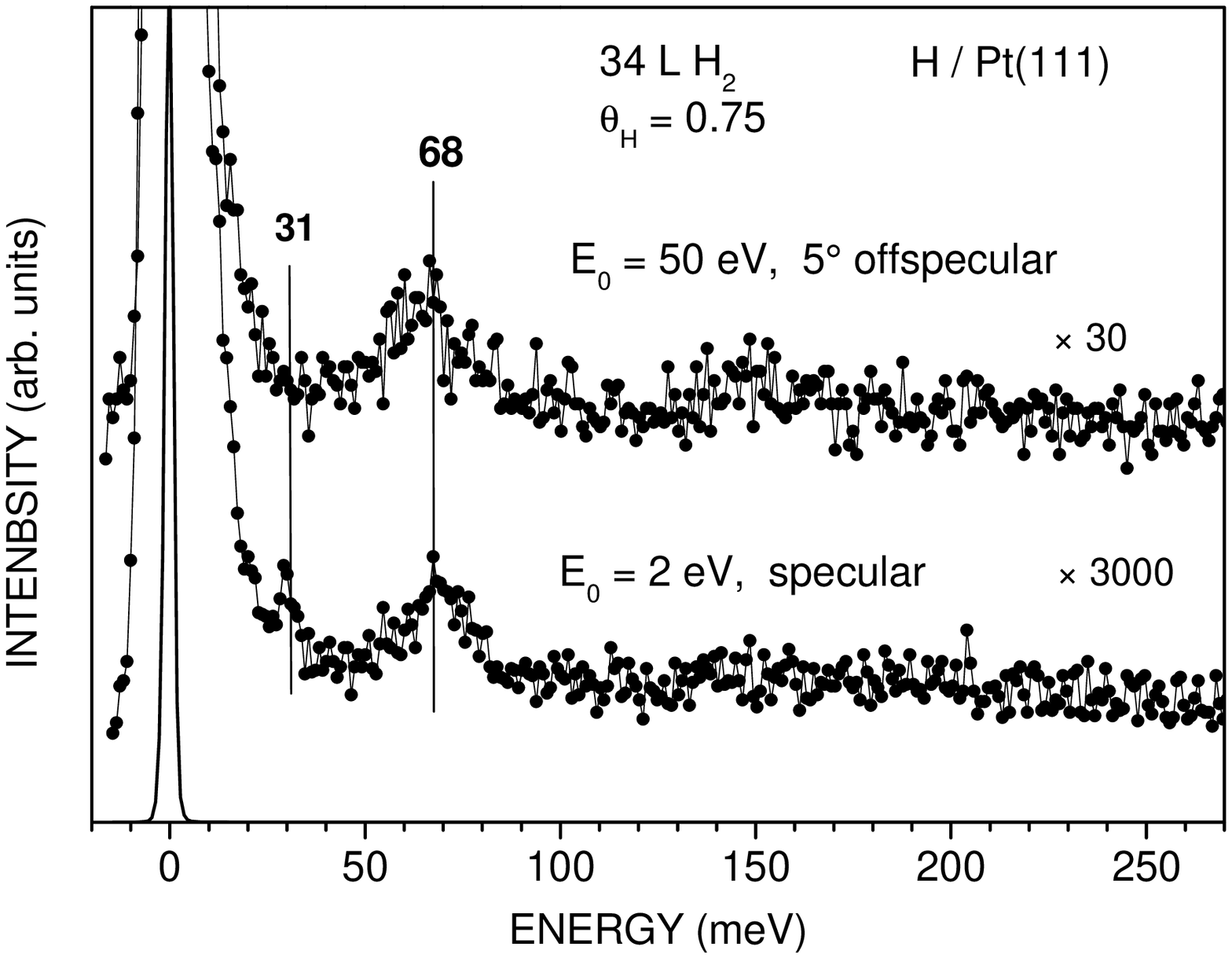}
\end{picture}
\caption{Comparison of HREEL spectra for H on Pt(111) for specular
and off-specular geometries} \label{HREELS2}
\end{center}
\end{figure}

The spectra for H and D for $\theta \le 0.75$ for specular
geometry are shown in Fig.\ref{HREELS1}. It is well known that the
loss intensity for vibrations of atomic H on transition metal
surfaces is quite low since it is mainly caused by impact
scattering. In spite of these intensity problems (note the
magnifications in Fig. \ref{HREELS1}) the spectra clearly indicate
the two H derived losses at  $31$ and $68$~meV. Note that these
new observations are quite different from previous results
obtained at a full ML \cite{Bar79,Ric87}. In particular, the
previously reported peaks at $112$~meV and $153$~meV are not
observed at these lower coverages. In Fig. \ref{HREELS2}, we also
show the results for off-specular geometry where the $31$ meV peak
disappears while the $68$ meV remains, indicating that the former
is mostly of dipole character. For deuterium, there is a clear
loss peak at $23.5$ meV, corresponding to the $31$ meV peak for
hydrogen with an isotope downshift. There is a very broad higher
mode between $40$ and $60$ meV not very well resolved from the
noise background.

We now turn to the theoretical calculations for the vibrational
excitations. The FP calculations for 3D APES were done
self-consistently with a parallel plane-wave code \cite{Laa93}
based on density functional theory with the GGA \cite{Per96}. The
Troullier-Martins pseudopotential \cite{Tro91} was used for Pt and
the Vanderbilt nonlocal ultrasoft pseudopotential \cite{Van90} for
H. Electronic wave functions were expanded in a plane-wave basis
with a kinetic energy cutoff at $35$~Ry. For the irreducible
Brillouin-zone integration eight special $\vec k$ points
\cite{Mon76} were used, together with a Fermi smearing of
$0.2$~eV. The calculated lattice constant for bulk Pt was
$a=7.568$~a.u. (experimentally $a=7.415$~a.u. \cite{Ash76}). The
supercell ($2\times 2\times 4$) has 16 Pt atoms and one H atom,
corresponding to a quarter monolayer coverage. The $z$ axis is
oriented towards the vacuum which was $14$~{\AA} thick. The
coordinates of the Pt atoms were relaxed before adding the H. Only
the vertical coordinate of the H atom was relaxed during the
calculations for each $(x,y)$ position of the APES.
We also checked the effects of the relaxation of Pt surface atoms.
We found that the potential energy differences between the fcc,
bridge and hcp sites changes less than 5~meV \cite{variation}.

\begin{figure}[htbp]
\unitlength1cm
\begin{center}
\begin{picture}(9.,7.)
\includegraphics{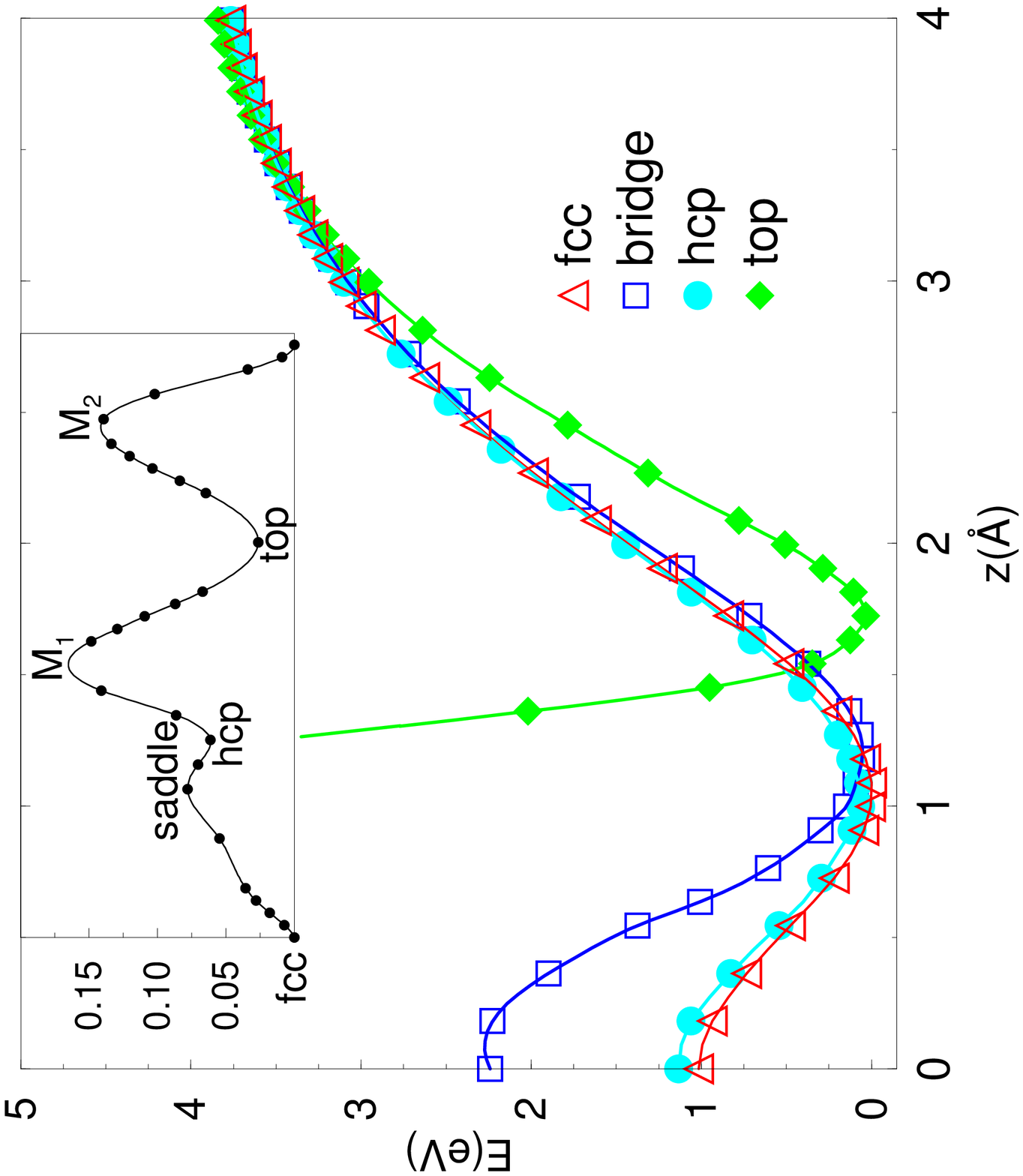}
\end{picture}
\caption{Vertical cuts in the 3D APES at high-symmetry sites. The
topmost atomic plane is at $z=0$, and the vacuum is at $z > 0$.
The inset shows the 2D minimum APES along the path
fcc-bridge-hcp-top-fcc.} \label{Corrugation}
\end{center}
\end{figure}

We first present results for the 3D APES as shown in
Fig.~\ref{Corrugation}. The 2D minimum APES is shown in the inset.
The top site is metastable and only $25$~meV above the lowest
energy adsorption fcc site and is surrounded by large energy
barriers. Along the fcc-hcp-fcc path, there is  a shallow barrier
of $78.9$~meV. At all sites the H~(D) is chemisorbed with an
electronic charge transfer from the substrate toward the adatom,
similar to that reported for H/Ni(111) \cite{Kre00}. These results
agrees well with other recent calculations
\cite{Pap00,Ols99,Wat01,Nob00,Kal02}. In particular,
the approach of K\"{a}ll\'{e}n and Wahnstr\"{o}m \cite{Kal02} is closest to the present
work and the resultant APES are very similar. However, there are
quantitative differences particularly concerning the curvature and
anharmonicity at the fcc sites. These will lead to substantial
differences for the vibrational band excitations as detailed
below.

\begin{figure}[htbp]
\unitlength1cm
\begin{center}
\begin{picture}(9.,6.)
\includegraphics{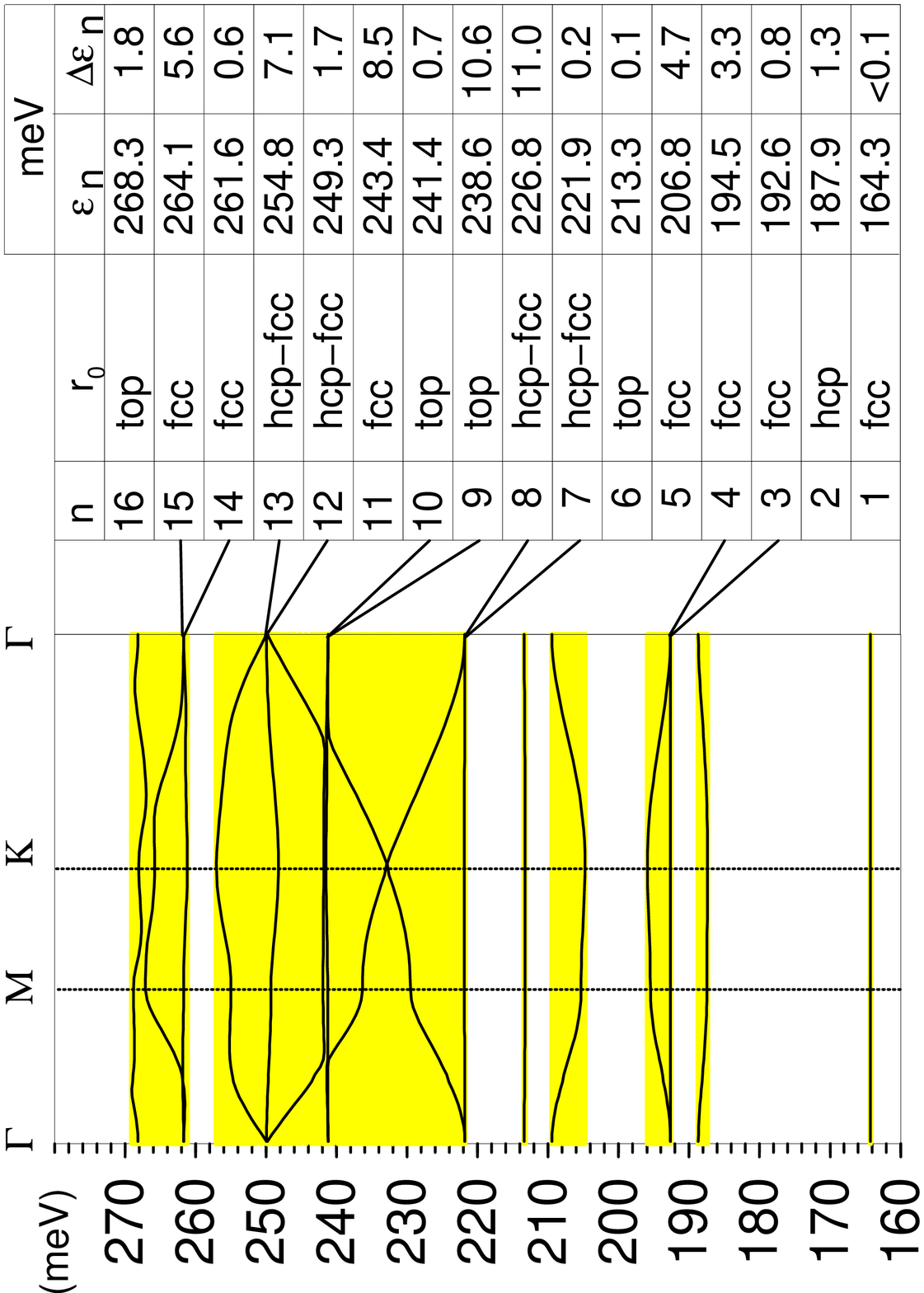}
\end{picture}
\caption{The lowest vibrational branches for H/Pt(111).
$\varepsilon_n$ denotes the band center, $\Delta \varepsilon_n$
the bandwidth, and $r_0$ the center of the BS with $\vec{k}=0$.}
\label{Branches1_16}
\end{center}
\end{figure}

To study the vibrational modes of the H and D adatoms, we
calculated the band states for the single particle  Hamiltonian
with the FP 3D APES. The Hamiltonian is diagonalized in  a basis
set $\{|\vec{G},n\rangle\}$ given by a direct product between
$595$ in-plane plane waves $\langle \vec{r}|\vec{G}\rangle$ and
$14$ harmonic-oscillator states $\langle z|\Psi_n\rangle$ for the
$z$ coordinate.
The lower branches $n=1-16$ for H are shown schematically in Fig.
\ref{Branches1_16}, along with the corresponding centers $\vec
{r}_0$ of the probability distributions for the $\vec k=0$ states,
the band centers $\varepsilon_n$, and the widths $\Delta
\varepsilon_n$. In addition, many modes have mixed in-plane and
vertical character and cannot be described in conventional AS or
SS terms. Also, the intervals between the band centers do not
match with the local harmonic oscillator model based on the
curvatures at the potential minimum.
 Finally, the higher
bands have considerable bandwidth, reflecting the delocalized
nature of the adsorbed H through tunnelling processes \cite{Chr79,Pus85}.

The first band $n=1$ localized at the fcc site is
  very narrow ($\Delta\varepsilon_1 < 0.1$~meV).
 The next band $n=2$ is localized at the hcp site. Its center is at $\varepsilon_2=23.6$~meV
 above the center of the lowest band and it has a bandwidth of $\Delta
\varepsilon_2=1.3$~meV. The difference $\varepsilon_2-\varepsilon_1$ confirms that even with the zero point
 energies taken into account, adsorption at low temperatures is
 confined to the fcc sites. To analyze the vibrational excitation
 energies arising from transitions starting from the lowest band $n=1$, we will
 focus only on those excited band states with finite amplitudes at the fcc
 site. The first excitations are toward the composite band $3\oplus
 4$, which has a total bandwidth of $4.1$ meV. The two subbands
 are degenerate at $\vec{k}=0$, with $E$ symmetry. The excitation to
 these bands, calculated as the difference between the bandcenters $\varepsilon_{3\oplus4}-\varepsilon_1$, occurs at $29.1$ meV. Taking into account the bandwidths,
 this is in excellent agreement with the new mode at $31$ meV observed in the present
 experiment. Moreover, this mode has mixed in-plane and
 vertical character with sizable dipole elements in the normal
 direction. This is confirmed by the fading  of the $31$ meV peak
 in the off-specular direction (see Fig.\ref{HREELS2}).

 The next  excitation  is to the  band $n=5$, and to the $7\oplus8$ pair of bands.
 The former is of $A$ symmetry has a bandwith of $4.7$ meV, while the
 latter has two subbands degenerate at $\vec{k}=0$ (with $E$ symmetry) and a total bandwidth of $11.2$~meV. The excitation energies are $42$~meV and $60$~meV,
 respectively. These bands have mainly an in-plane character. We suggest that the observed peak centered at
 $68$~meV could be a composite of these unresolved transitions. Another
 point to note about the higher bands $7\oplus8$ is that they have
 considerable weight at both the fcc and hcp sites. The states in
 these excited bands have high mobility throughout the surface. The
 excitation energy $60$~meV needed to reach these bands is close to
 the activation energy observed in the quasi-elastic Helium Scattering
 experiment by Graham {\it et al.} \cite{Gra99}. Thus, our band state results
support the picture that the diffusion of H on Pt(111) happens not through classical
hopping but rather via the activated tunnelling mechanism as
recently proposed \cite{Bad01} to account for the similar behavior
observed for H/Ni(111). The results give a direct confirmation of the protonic band picture
 introduced in \cite{Chr79,Pus85}.

When D is substituted for H in the calculations, the lower bands
 have a one to one correspondence with the H case with a
downshift of frequency approximately by a factor of $1/\sqrt{2}$
as expected. For deuterium, the first  excitation energy is at
$23.2$~meV toward a composite band of width $0.2$~meV, in excellent agreement with the experimental
observation for D as shown in Fig.\ref{HREELS1}. There are
transitions to higher bands at excitation energies between $30$ and
 $60$~meV: an excitation to an $A$ band at $30.8$~meV (bandwidth $0.8$~meV), one to
 an $E$ composite band at $46$~meV (bandwidth $3.4$~meV), and another to an $E$
 composite band at $59.8$~meV (bandwidth $2.9$~meV). These could account for the broad peak centered at
$49$ meV observed in the experiment.

There are many more excitations to the higher bands that we have
not addressed here. Experimentally, they are not observed at low
coverages ($\theta \leq 0.75$ ML).
 Qualitatively, one can understand
this by noting that the  higher bands are much more extended than
the ground states and thus the matrix elements should be much
smaller. As such, the resultant weak and broad peaks are hard to
observe. The picture changes at full ML because then the
tunnelling of the H atom would be suppressed by on-site exclusion
and effectively the H atoms are localized again. Interestingly, at
a full ML, the $31$~meV peak disappears from the observations and two
additional peaks appear at $110$~meV and $153$~meV \cite{Jac01} in
the present experiment. These loss peaks agree nicely with the
earlier results of Richter and Ho \cite{Ric87}.

In summary, we have presented a combination of theoretical
calculations and experiments on the energetics and vibrational
properties of H and D adatoms on Pt(111). The experimental data
for the loss peaks at low coverages ($\theta \leq 0.75$ ML) differ
significantly from previous studies at full H coverage. There is
good agreement between the observed HREELS loss peaks and the
protonic band structure calculations derived from a full 3D APES.
This demonstrates that the vibrational spectroscopy of H adsorbed
on metals and in particular for H/Pt(111) cannot be understood
simply in terms of localized AS and SS oscillation modes at the
adsorption sites, as often assumed in the literature. This picture
is also crucial for a proper understanding of the recent diffusion
data for   H/Pt(111) \cite{Gra99}. Finally, we emphasize that the
dynamical behavior of this system at full ML is quite different
from that at low coverages. There is also a possibility of change
of adsorption sites. Further studies for the full ML case are in
progress.

Acknowledgements: This work has been in part supported by the
Academy of Finland through its Center of Excellence program and in
part by a Galkin Fund Fellowship. We
acknowledge generous computing resources from CSC -- Scientific
Computing Ltd., Espoo, Finland. Discussions with K. Honkala, K. E.
Laasonen, M. J. Puska, R. M. Nieminen,  and O. S. Trushin are also greatly appreciated. We thank
J. P. Toennies for bringing our groups in contact.

\begin{thebibliography}{200}
\bibitem{peaks} The spectra at higher coverages contain HREELS peaks
that are not visible for low coverages ($\theta \in [0.33,0.75]$).
This may be due either to a small intensity which can hide the
modes in the background noise or to a domination of interactions
between adsorbates. The high coverage case will be analyzed in a
separate publication \cite{Jac01}.
\bibitem{variation}We have explicitly checked that  the transition
energies are not sensitive to small variations in the APES due to
the relaxation of surface atoms and to the interpolation scheme.
\end{thebibliography}
\end{document}